\documentclass[12pt]{iopart}

\usepackage{iopams}
\usepackage{eqnarray} 
\usepackage{graphicx}
\usepackage{color}
\usepackage{cite}
\usepackage[colorlinks=true,linkcolor=blue,citecolor=red]{hyperref}
\newcommand{\be}{\begin{equation}}
\newcommand{\ee}{\end{equation}}
\newcommand{\bea}{\begin{eqnarray}}
\newcommand{\eea}{\end{eqnarray}}

\begin{document}

\title[BDD and ladder operators]{Ladder operators for the Ben Daniel-Duke Hamiltonians and their SUSY partners}

\author{M.I. Estrada-Delgado$^1$ and David J. Fern\'andez$^2$}

\address{Departamento de F\'isica, Cinvestav, A.P. 14-740, 07000 Ciudad de M\'exico, M\'exico}
\ead{$^1$iestrada@fis.cinvestav.mx ,$^2$david@fis.cinvestav.mx}
\vspace{10pt}
\begin{indented}
\item[]February  2019
\end{indented}

\begin{abstract}
Position dependent mass systems can be described by a class of operators which include the Ben Daniel-Duke Hamiltonians. The usual methods to solve this kind of problems are, in general, either numerical or those looking for a connection with constant mass problems. In this paper we impose the existence of first-order ladder operators to fix our initial system. Then, we perform the first and second order supersymmetric transformations to generate families of Hamiltonians whose eigenfunctions are known analytically for a given mass profile.
\end{abstract}

%%%%%%%%%%%%%%%%%%%%%%%%%%%%%%%%%%%%%%%%%%%%%%%%%%%%%%%%%%
\section{Introduction}

The description of charge carriers inside semiconductor heterostructures has been described historically by position dependent mass (PDM)  Hamiltonians, whose deduction from first principles produces a wide family with different properties, ranging from non Hermitian to Hermitian operators with several types of boundary condition\cite{Von1983,Li1993,Gor1969}.

On the other hand, the generation of solvable systems is of practical interest, although it is also important theoretically since the identification of any new solvable Hamiltonian opens the door to unexplored areas and supplies a bunch  of new information to be analyzed. In the literature the solution to this kind of problems usually involves either  numerical methods or to look for a connection between a PDM system and a constant mass solvable Hamiltonian \cite{Koc2010,Cru2008}.

In quantum mechanics one of the pillar examples is the harmonic oscillator, since it appears naturally in many areas of physics and can be solved in several ways, one of which is the algebraic method where the Hamiltonian and the ladder operators satisfy the so called Heisenberg-Weyl algebra. This algebraic treatment makes the Hamiltonian to be factorized also by the ladder operators, while the commutator between the last becomes a constant \cite{Zet2009}. The modification of this scheme has been the key for arriving to novel important topics, like the polynomial Heisenberg algebras (PHA) of degree $m$ where the commutator between the new ladder operators is no longer a constant but it is an $m$ degree polynomial in the Hamiltonian  \cite{Fer2004}, or the generalized factorization, where the intertwining operators factorize the harmonic oscillator Hamiltonian just in a given order but in the opposite direction produces a Hamiltonian different from the harmonic oscillator one \cite{Mie1984,Fer1984,Ros1998}. As pointed out in \cite{Per1996}, however, the ladder operators do not necessarily factorize the involved Hamiltonian, as it happens for example with the PHA of degrees greater than zero. Keeping this in mind, we are going to seek the analogues of the PHA for position dependent mass system. As a first step of this approach, we will look for those PDM systems that posses first-order ladder operators, and we will analyze the associated potential and ladder operators which are involved. Since this simple example is connected with the zero degree PHA, it turns out that the corresponding ladder operators factorize also the Hamiltonian involved, thus it should be related to the work of Cruz, Negro and Nieto \cite{Cru2008}.
 
%%%%%%%%%%%%%%%%%%%%%%%%%%%%%%%%%%%%%%%%%%%%%%%%%%%%%%%%%%%%%%%%
For constant mass problems Witten's supersymmetric quantum mechanics (SUSY QM) has proven to be a powerful tool for generating new solvable Hamiltonians departing from a given initial one \cite{Wit1981,Fer1998,Ber2014,Fer2014}. One would expect that the technique could be also applied in our case, with the initial system being automatically fixed once the ladder operators are required to exist. Although the use of SUSY quantum mechanics for PDM systems is not new \cite{Nie2007,Koc2003}, the main difference here is that we will deal with the initial and final Hamiltonians in their original form, without trying to establish any connection with a constant mass problem (see also \cite{Sch2006}).

Let us stress once again that this article is just the first step toward a wider goal, the search of general PDM Hamiltonians ruled by PHA and the analysis of their SUSY partners. The cases for degrees greater than zero will be addressed in subsequent works.

In the next section we will discuss the PDM system we are interested in, described by the Ben Daniel-Duke (BDD) Hamiltonian. Then, in section 3 we will determine the general form of the associated potential, for systems which have first-order differential ladder operators and give place to realizations of the zero degree PHA. Section 4 contains some examples of such a general family of potentials for different mass profiles. In section 5 we will quickly review the standard SUSY QM for constant mass Hamiltonians. Section 6 reports how to apply the first and second-order SUSY methods to the BDD Hamiltonian. In section 7 we will employ these SUSY results for the cosine mass profile. Section 8 contains our conclusions.

%%%%%%%%%%%%%%%%%%%%%%%%%%%%%%%%%%%%%%%%%%%%%%%%%%%%%%%%%%
\section{Ben Daniel-Duke and von Roos Hamiltonians}

As pointed out in \cite{Von1983,Li1993}, there are several ways to define the kinetic term for a PDM Hamiltonian. One of the most general, which is automatically Hermitian, was proposed by von Roos in 1993 in the form:
\be
H_{VR}=\frac{1}{4}(m^{\alpha}\hat{p} m^{\beta}\hat{p} m^{\gamma}+m^{\gamma}\hat{p} m^{\beta}\hat{p} m^{\alpha})+V,
\label{VR1}
\ee
%\noindent
where $m$ depends on the position, $V$ is a real potential and $\alpha,\beta,\gamma$ are constants subject to the constrain $\alpha+\beta+\gamma=-1$. Different selections of these constants lead to different Hamiltonians, and a given choice is based typically on physical considerations\cite{Sou2006}. However, the problem can be addressed from a different point of view by writing (\ref{VR1}) as the differential operator:
\be
H_{VR}=-\frac{\hbar^{2}}{2m(x)}\frac{d^{2}}{dx^{2}}+\frac{\hbar^{2}m'(x)}{2m^{2}(x)}\frac{d}{dx}+V_{eff}(x)
\label{VR2},
\ee
with
\be
\hskip-1cm V_{eff}(x)=V(x)+\hbar^{2}\left(\frac{(1+\beta)m''(x)}{4m^{2}(x)}-\frac{[m'(x)]^{2}}{2m^{3}(x)}(\alpha^{2}+\alpha\beta+\alpha+\beta+1)\right).
\ee

In the subsequent treatment, instead of equation (\ref{VR1}) we will work with expression (\ref{VR2}), which involves $V_{eff}$. If one is interested in one particular ordering defined by equation (\ref{VR1}), $V(x)$ has to be expressed in terms of $V_{eff}$ and an extra term arising from such ordering, i.e.,
\be
\hskip-1cm V(x)=V_{eff}(x)-\hbar^{2}\left(\frac{(1+\beta)m''(x)}{4m^{2}(x)}-\frac{[m'(x)]^{2}}{2m^{3}(x)}(\alpha^{2}+\alpha\beta+\alpha+\beta+1)\right).
\ee
Let us note also that expression (\ref{VR2}) has the form proposed by Ben Daniel and Duke in 1996 \cite{Ben1996}:
\bea
\hskip-1cm H_{BDD}&=\frac{1}{2}\left(\hat{p}\frac{1}{m(x)} \hat{p}\right)+V_{BDD}
= -\frac{\hbar^{2}}{2m(x)}\frac{d^{2}}{dx^{2}}+\frac{\hbar^{2}m'(x)}{2m^{2}(x)}\frac{d}{dx}+V_{BDD}(x). \label{BDD}
\eea

\noindent
From now on we will work with the BDD Hamiltonian (\ref{BDD}) and the corresponding subscripts will be dropped.

%%%%%%%%%%%%%%%%%%%%%%%%%%%%%%%%%%%%%%%%%%%%%%%%%%%%%%%%%%
\section{General form of the BDD Hamiltonian with first order ladder operators}

Let us consider a system ruled by a BDD Hamiltonian $H$, as given in equation (\ref{BDD}), and two ladder operators $L^{\pm}$ such that the following commutation relations hold,
\be
[H,L^{\pm}]=\pm \Delta_E L^{\pm},\hspace{1cm} \Delta_E \in\mathbb{R},
\label{3.6}
\ee
thus imitating the harmonic oscillator algebra. Equation (\ref{3.6}) ensures that, given an eigenfunction $\psi$ of $H$ with eigenvalue $\epsilon$, $H\psi=\epsilon\psi$, the action of the ladder operators $L^\pm$ onto such a $\psi$ produces new eigenfunctions of $H$ with eigenvalues $\epsilon\pm \Delta_E$, as long as they satisfy the boundary conditions, i.e.,
\be
H(L^{\pm}\psi)=(\epsilon\pm \Delta_E)(L^{\pm}\psi).
\ee 
Since the BDD Hamiltonian is Hermitian, we can choose $L^{\pm}$ as being adjoint to each other, which for non-Hermitian Hamiltonians can not be done in general.

If the energy spectrum of $H$ is to be bounded from below, there should exist a set of formal eigenfunctions of $H$ which also would be annihilated by $L^{-}$, $L^{-}\psi_{0}=0$. Some of them could fulfill the boundary conditions, thus the corresponding eigenvalues will belong to the spectrum of $H$ and they will be called extremal states of the system, in analogy with the constant mass problems ruled by polynomial Heisenberg algebras. The lowest energy level associated to those extremal states will be called also ground state energy.

Let us assume now that $L^{-}$ is a first order differential ladder operator of the form:
\be
L^{-}=\frac{1}{\sqrt{2}}\left(\alpha _1(x)\frac{d}{dx}+\beta _1(x)\right).
\ee
The requirement that equation (\ref{3.6}) should be fulfilled leads to a set of coupled differential equations relating $m(x)$, $\alpha_{1}(x)$, $\beta_{1}(x)$ and $V(x)$, namely,
\bea
&& \hskip-1.5cm \alpha _1'(x)+\frac{m'(x)}{2 m(x)}\alpha _1(x)=0, \\
&& \hskip-1.5cm \beta _1'(x)=-\frac{1}{2}\alpha _1''(x)+\frac{m'(x)\alpha_{1}'(x)}{2m(x)}
+\alpha _1(x)\left[\left(\frac{m'(x)}{m(x)}\right)^{2}+\frac{\Delta_E m(x)}{\hbar^{2}}-\frac{m''(x)}{2m(x)}\right],
\nonumber \\
&& \hskip-1.5cm \alpha _1(x) V'(x)=\Delta_E \beta _1(x)+\frac{\hbar^{2}m'(x) \beta _1'(x)}{2 m(x)^2}-\frac{\hbar^{2}\beta _1''(x)}{2 m(x)}.
\eea
Since the mass profile is to be fixed by physical reasons, it is natural to proceed by solving this set of equations in terms of $m(x)$. The corresponding solution is given by:
\bea
&& \hskip-1.0cm \alpha _1(x)=a m(x)^{-\frac{1}{2}}, \\
&& \hskip-1.0cm \beta _1(x)=\frac{a}{2} \left(m(x)^{-\frac{1}{2}}\right)'+\frac{a \Delta_E }{\hbar^2} \int m(x)^{\frac{1}{2}}dx, \\
&& \hskip-1.0cm V(x)=\frac{1}{2}\left(\frac{\Delta_E}{\hbar}\right)^2 \left[\int m(x)^{\frac{1}{2}}dx\right]^2-\frac{\hbar^2}{8} \left(\left(m(x)^{-\frac{1}{2}}\right)'\right)^2-\frac{\hbar^2\left(m(x)^{-\frac{1}{2}}\right)''}{4 m(x)^{\frac{1}{2}}}.
\eea
As $L^{-}$ is a first order differential operator, we just get one formal eigenfunction for the extremal states of the system, which is given by:
\be
\psi _0(x)=c_{0} \, m(x)^{\frac{1}{4}}\exp \left(-\frac{\Delta_E}{2\hbar^{2}}\left[\int m^{\frac{1}{2}}dx\right]^2\right),
\ee 
and it has associated the ``energy''
\be
E_{0}=\frac{\Delta_E}{2}.
\ee
As it was said previously, in this case the adjoint of the annihilation operator is the creation operator, which is given by:
\be
L^{+}=\frac{1}{\sqrt{2}}\left(-\alpha _1(x)\frac{d}{dx}+\beta _1(x)-\alpha_{1}'(x)\right).
\ee 
Thus, the excited state candidates will be generated from the iterated action of $L^{+}$ onto the ground state $\psi_{0}$, namely, $\psi_{n}(x) \propto (L^{+})^{n}\psi_{0}$, whose energies are $E_{n}=E_{0}+n\Delta_E$, $n\in\mathbb{N}_{0}$. 

Once the ladder operators have been set, it is straightforward to calculate the commutator between them, which is given by
\be
[L^-,L^+]=\frac{a^2\Delta_E}{\hbar^2} .
\ee

Along this work we will restrict ourselves to cases where the oscillation theorem is valid. Thus the presence or absence of the corresponding formal eigenvalues in the spectrum of $H$ will depend on either the generated states satisfy or not the boundary conditions imposed by Sturm-Liouville theory.

%%%%%%%%%%%%%%%%%%%%%%%%%%%%%%%%%%%%%%%%%
\subsection{Boundary conditions}
\label{bounc}

From the point of view of Sturm-Liouville theory, the operator (\ref{BDD}) is \emph{formally self adjoint} \cite{Alg2008}. At this stage, equation (\ref{BDD}) is not enough to ensure that the system has real eigenvalues whose eigenfunctions $y(x)$ form an orthogonal basis of ${\cal L}^{2}$ in a domain $(x_0,x_1)$. In fact, in the previous algebraic treatment one could have generated solutions that are not orthogonal to each other, thus making $H$ not being self adjoint in the space generated by them. However, by imposing on two generated solutions $y_1, \ y_2$ the boundary conditions
\be
p(y_1'\bar{y_2}-y_1\bar{y_2}')|_{x_0}^{x_1}=0,\hspace{1cm}p=\frac{1}{m}, \hspace{1cm}
\label{bcsl}
\ee
a self adjoint Hamiltonian $H$ is guaranteed, and an orthogonal basis of ${\cal L}^{2}$ is ensured.

Notice that for an eigenfunction $y(x)$ equation (\ref{bcsl}) is equivalent to 
\bea
c_{1}y(x_0)+c_{2}p(x_0)y'(x_0)=0, \nonumber
\\
c_{1}y(x_1)+c_{2}p(x_1)y'(x_1)=0.
\label{bcsleq}
\eea
In case that either $x_0$ or $x_1$ is infinite, equations (\ref{bcsl}) and (\ref{bcsleq}) must be interpreted as a limit.
%%%%%%%%%%%%%%%%%%%%%%%%%%%%%%%%%%%%%%%%%
\section{Examples of PDM systems with first order ladder operators}

It is not hard to find in the literature examples where structures of $GaAs$ and $Al_{z}Ga_{1-z}As$ are created \cite{Li1993}. Inside this type of semiconductors the electrons and holes posses an effective mass that varies according to $m=m_0+m_1 z$, $m_0$ and $m_1$ being two constants defined by the material. The variable $z$ is defined by the gradient of $Al$ in the structure; specifically, in \cite{Li1993} $z$ varies as an error function, but also it is suggested that the concentration can adopt somehow any desired profile. This suggestion was used in \cite{Koc2005,Kho2011}, where quadratic inverse and linear inverse profiles were considered. Below we will stick as well to this proposal, by choosing simple profiles that will supply us physically interesting information.

%%%%%%%%%%%%%%%%%%%%%%%%%%%%%%%%%%%%%%%
\subsection{Quadratic profile}

Working in atomic units, let us assume that the mass profile takes the form $m(x)=\frac{x^{2}}{2}+m_0$. Simple and straightforward calculations lead us to the following expressions, where the potential has an infinite equidistant spectrum:
\bea
&& \hskip-2.5cm \alpha_{1}(x) =  \frac{a}{\sqrt{m_0+\frac{x^2}{2}}}, \\
&& \hskip-2.5cm \beta_{1}(x) =  \frac{a \left(\sqrt{2 m_0+x^2} \left(\frac{\Delta_E x}{2}-\frac{x}{\left(2 m_0+x^2\right)^2}\right)+m_0 \Delta_E \log \left(\sqrt{2 m_0+x^2}+x\right)\right)}{\sqrt{2}},\\
&& \hskip-2.5cm V(x) =  \frac{1}{8} m_0 \Delta_E^2 x^2 +\frac{\Delta_E^2 x^4}{16}-\frac{5}{4 \left(2 m_0+x^2\right)^2} +\frac{7 m_0}{2 \left(2 m_0+x^2\right)^3}+ \frac{1}{4} m_0^2 \Delta_E^2 \log ^2\left(\sqrt{2 m_0+x^2}+x\right) \nonumber \\
&&  \hskip-1.2cm  +\frac{1}{4} m_0 \Delta_E^2 x \sqrt{2 m_0+x^2} \log \left(\sqrt{2 m_0+x^2}+x\right),
\eea
with a ground state wavefunction given by
\bea
\psi_{0}(x)=&\sqrt[4]{2 m_0+x^2} \left(\sqrt{2 m_0+x^2}+x\right)^{-\frac{1}{4} m_0 \Delta_E x \sqrt{2 m_0+x^2}}\times \nonumber 
\\
&\exp \left\{-\frac{1}{16} \Delta_E \left[4 m_0^2 \log ^2\left(\sqrt{2 m_0+x^2}+x\right)+2 m_0 x^2+x^4\right]\right\}.
\eea

\begin{figure}[htp]
\centering
\includegraphics[height=7cm, width=12cm]{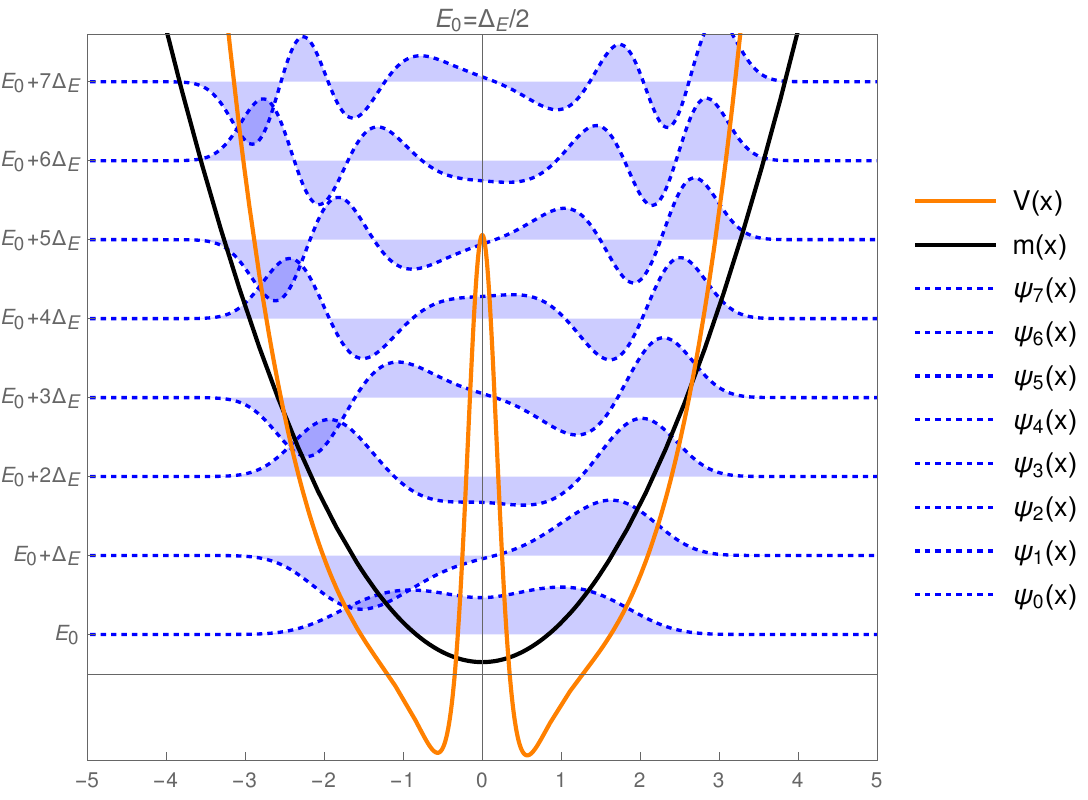}
\caption{Graphs for the quadratic mass profile with $m_0=0.15$, the corresponding potential with an infinite equidistant spectrum, and some eigenfunctions associated to the lowest eigenvalues in atomic units.}
\label{fig1}
\end{figure}

It can be seen clearly in Figure \ref{fig1} that around $x=0$ the shape of the potential changes drastically; however, it is non-singular for all $x\in\mathbb{R}$.

%%%%%%%%%%%%%%%%%%%%%%%%%%%%%%%%%%%%%
\subsection{Cosine mass profile}
\label{cosproex}

\begin{figure}[htp]
\centering
\includegraphics[height=7cm, width=12cm]{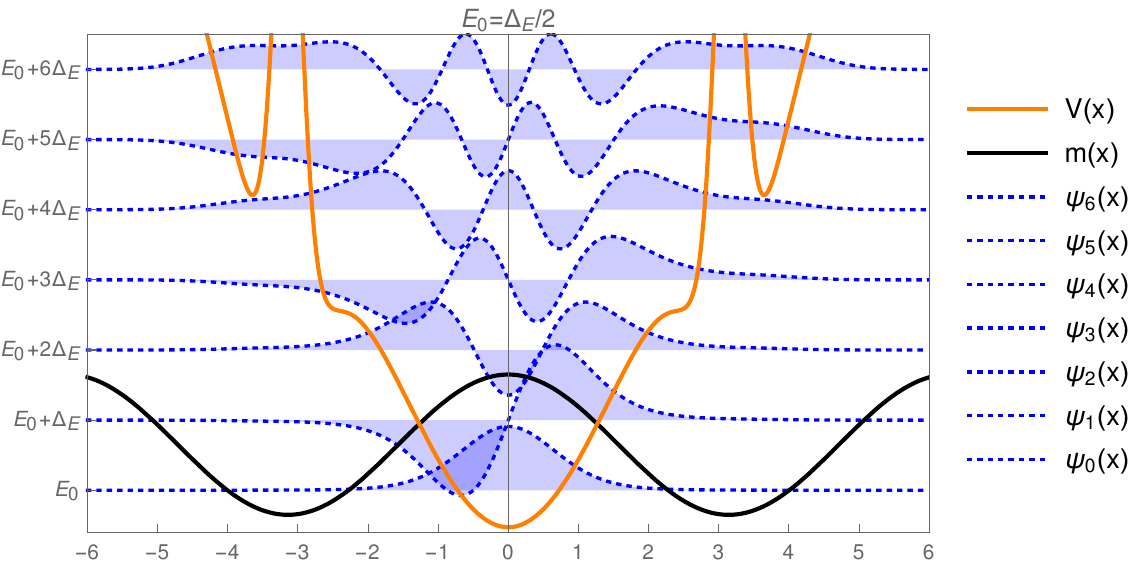}
\caption{Graphs for the cosine mass profile with $m_0=1.15$, the corresponding potential with an infinite equidistant spectrum, and some eigenfunctions associated to the lowest eigenvalues in atomic units.}
\label{fig2}
\end{figure}

Let us suppose now that a cosine mass profile $m(x)=\cos(x)+m_0$ of height $m_0>1$ is chosen. When $m_0$ is close to 1 the mass profile approaches to zero at the points $\pm(2n+1)\pi$, $n\in\mathbb{Z}$, and thus the effects on the potential induced by the proximity to a singularity look similar as the behavior around $x=0$ of the previous example (see figure \ref{fig2}). The relevant functions for the system are now explicitly given by:
\bea
&& \hskip-2cm \alpha_{1}(x) =  \frac{a}{\sqrt{m_0+\cos (x)}}, \\
&& \hskip-2cm \beta_{1}(x) = \frac{a \left(8 (m_0+1)^2 \Delta_E \left(\frac{m_0+\cos (x)}{m_0+1}\right)^{3/2} E\left[\frac{x}{2},\frac{2}{m_0+1}\right]+\sin (x)\right)}{4 (m_0+\cos (x))^{3/2}}, \\
&& \hskip-2cm V(x) = \frac{1}{64} \left(128 \Delta_E^2 (m_0+1) E\left[\frac{x}{2},\frac{2}{m_0+1}\right]^2+\frac{-8 m_0 \cos (x)+3 \cos (2 x)-11}{(m_0+\cos (x))^3}\right),
\label{Vcos}
\eea
while the ground state eigenfunction reads
\be
\psi_{0}=\sqrt[4]{m_0+\cos (x)} e^{-2 \Delta_E (m_0+1) E\left[\frac{x}{2},\frac{2}{m_0+1}\right]^2}.
\ee
In the previous expressions $E[\phi,k]$ is the incomplete elliptic integral of second kind, defined by $E[\phi,k]=\int_{0}^{\phi}\sqrt{1-k\sin^{2}{\theta}}\ d\theta$.

\begin{figure}[htp]
\centering
\includegraphics[height=7cm, width=12cm]{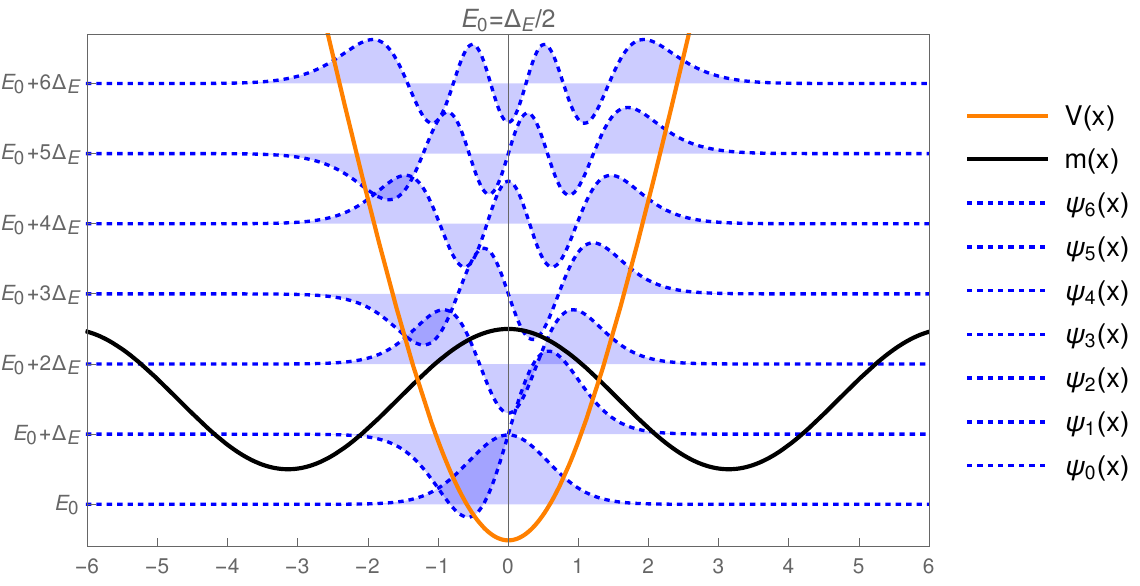}
\caption{The same system as in Figure \ref{fig2} but for $m_0=2$. For low energies it resembles the harmonic oscillator almost perfectly.}
\label{fig3}
\end{figure}

Let us note that when the mass profile changes slowly, or it is high enough to look like if it would have this property, the potential tends to the harmonic oscillator, as it can be seen in Figure \ref{fig3}.

%%%%%%%%%%%%%%%%%%%%%%%%%%%%%%%%%%%%%%%%%%%%%%%%%%
\subsection{Linear profile}
As our last example, let us assume that the mass profile varies linearly as $m(x)=x$, which will induce a singularity in the potential at $x=0$. The functions defining the ladder operator and the potential are given now by:
\bea
\alpha_{1}(x) & = & \frac{a}{\sqrt{x}}, \\
\beta_{1}(x) & = & \frac{1}{8} a \left(\frac{16}{3} \Delta_E x^{3/2}-\frac{2}{x^{3/2}}\right), \\
V(x) & = & \frac{1}{96} \left(\frac{64 \Delta_E^2 x^3}{3}-\frac{21}{x^3}\right),
\eea
with an extremal state given by
\be
\psi_{0}(x)=\sqrt[4]{x} e^{-\frac{1}{9} \left(2 \Delta_E x^3\right)}.
\ee

\begin{figure}[htp]
\centering
\includegraphics[height=7cm, width=10cm]{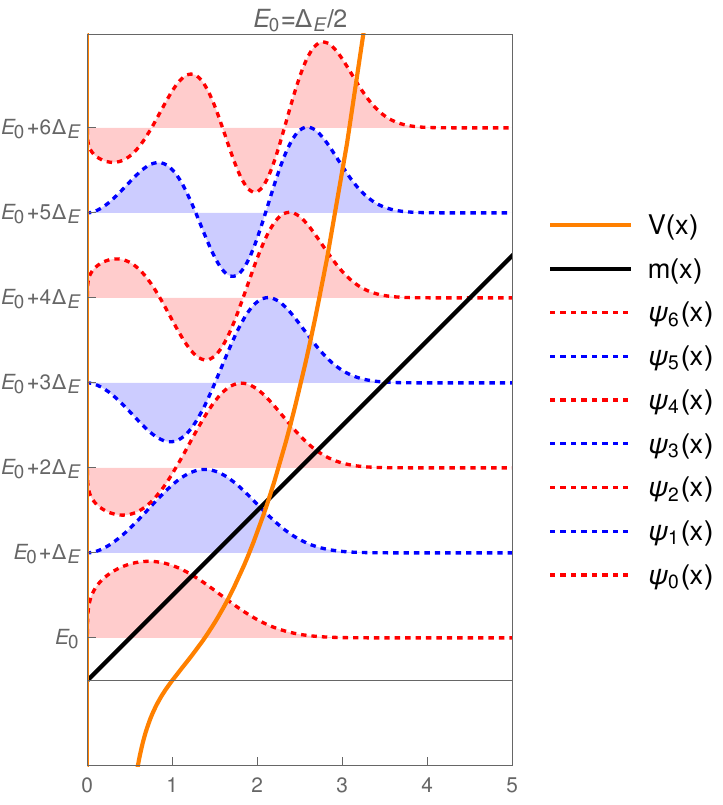}
\caption{Graphs for the linear mass profile, the corresponding potential with an infinite equidistant spectrum, and some eigenfunctions associated to the lowest formal eigenvalues in atomic units.}
\label{fig4}
\end{figure}

In Figure \ref{fig4} we can see that, among all the formal eigenfunctions generated by acting iteratively $L^{+}$ onto $\psi_{0}$, only half of them satisfy the boundary conditions at the origin discussed in section \ref{bounc} (those labeled by odd indexes).

%%%%%%%%%%%%%%%%%%%%%%%%%%%%%%%%%%%%%
\section{Standard first order SUSY transformation}

In the first order supersymmetric quantum mechanics for constant mass problems, there are two first-order intertwining operators and two hermitian Hamiltonians of Schr\"{o}dinger type, one of them supposed to be exactly solvable, which are related by:
\bea
&\tilde H A^{\dagger}=A^{\dagger}H,  \qquad HA=A\tilde H.
\eea
This ensures that the eigenfunctions of one Hamiltonian are mapped into those of the other one and vice versa, with the possible exception of one eigenstate. In general, the mapped solutions do not necessarily satisfy the given boundary conditions.

Let $\psi$ be an eigenfunction of $H$ with eigenvalue $\epsilon$, and consider as well an eigenfunction $\tilde{\psi}$ of $\tilde{H}$ with eigenvalue $\tilde{\epsilon}$. The action of $A^{\dagger}$ on the first one, or $A$ on the second, generates a solution of the corresponding SUSY partner Hamiltonian, namely,

\bea
\tilde H(A^{\dagger}\psi)=A^{\dagger}( H\psi)= \varepsilon(A^{\dagger}\psi),
\\
H(A\tilde\psi)=A(\tilde H\tilde\psi)=\tilde\varepsilon(A\tilde\psi).
\nonumber
\eea

Now by choosing one of the formal eigenfunctions $u$ of the initial Hamiltonian $H$, $Hu=\epsilon u$, although it does not necessarily satisfy the boundary conditions of the problem, the intertwining operator $A^{\dagger}$ takes the form:
\be
A^{\dagger}=\frac{1}{\sqrt{2}}\left(-\frac{d}{dx}+W\right),\hspace{1cm}W=\frac{u'}{u}.
\ee
The function $W$ is called {\it superpotential} while $u$ is the {\it seed solution}, and both relate the potentials of the initial and new Hamiltonians as follows:
\be
\tilde V=V-W'=V-\frac{d^{2}\ln{u}}{dx^{2}}.
\ee
Moreover, it is straightforward to show that one formal eigenfunction of $\tilde H$ associated to $\epsilon$ exists, which is given by $1/u$, i.e,
\be
\tilde H\left(1/u\right)=\epsilon \left(1/u\right).
\ee

%%%%%%%%%%%%%%%%%%%%%%%%%%%%%%%%%%%%%%%%%%
\section{Position dependent mass SUSY transformations}

\subsection{First-order transformations}

As mentioned previously, the use of SUSY quantum mechanics to generate solutions of position dependent mass systems is not completely new (see for example \cite{Nie2007,Sch2006}). In order to implement such a method, let us consider two real functions $m(x)$, $W_1(x)$, which define the first-order differential operator
\be
A_1=\frac{\hbar}{\sqrt{2m(x)}}\frac{d}{dx}+W_1(x),
\ee
and its adjoint 
\be
A_1^{+}=-\frac{\hbar}{\sqrt{2m(x)}}\frac{d}{dx}+W_1(x)+\frac{\hbar}{2\sqrt{2}}\frac{m'(x)}{(m(x))^\frac{3}{2}}.
\ee

The product of $A_1$, $A_1^{\dagger}$ in one given order leads to (for simplicity we do not write explicitly the $x$ dependence of the functions $m$, $W_1$, $V_0$ and $V_1$), 
\be
A_1^{\dagger}A_1=-\frac{\hbar^{2}}{2m}\frac{d^{2}}{dx^{2}}+\frac{\hbar^{2} m'}{2m^{2}}\frac{d}{dx}+W_1^{2}-\frac{\hbar W_1'}{\sqrt{2m}}+\frac{\hbar W_1 m'}{2\sqrt{2}m^{\frac{3}{2}}},
\label{RE}
\ee
while in the reverse order produces
\bea
\label{IE}
\hskip-1.8cm A_1A_1^{\dagger}=&-\frac{\hbar^{2}}{2m}\frac{d^{2}}{dx^{2}}+\frac{\hbar^{2} m'}{2m^{2}}\frac{d}{dx}+W_1^{2}+\frac{\hbar W_1'}{\sqrt{2m}}+\frac{\hbar W_1 m'}{2\sqrt{2}m^{\frac{3}{2}}} +\frac{\hbar^{2}m''}{4m^{2}}-\frac{3\hbar^{2}(m')^{2}}{8m^{3}}.
\eea
Since the kinetic part of the BDD Hamiltonian  (\ref{BDD}) is equal to the first two terms of the operators in equation (\ref{RE}) and (\ref{IE}), it is natural to identify,
\bea
\hskip-1cm H_{0}-\epsilon_{1}=A_1^{\dagger}A_1, \hspace{.3cm}H_{1}-\epsilon_{1}=A_1A_1^{\dagger}\hspace{.1cm}\Rightarrow\hspace{.1cm}H_{0}A_1^{\dagger}=A_1^{\dagger}H_{1},\hspace{.3cm} H_{1}A_1=A_1H_{0},
\label{Inter1}
\eea
while the corresponding potentials are expressed as
\bea
V_{0}& = & W_1^{2}-\left(\frac{\hbar W_1}{\sqrt{2m}}\right)'+\epsilon_{1},
\label{susyv0} \\
V_{1}& =& V_{0}+\frac{2\hbar W_1'}{\sqrt{2m}}-\frac{\hbar^{2}}{2\sqrt{m}}\left(\frac{1}{\sqrt{m}}\right)''.
\label{susyv1}
\eea

Given the initial potential $V_{0}$ and the mass profile $m$, the Riccati equation (\ref{susyv0}) has to be solved to find the analogue of the superpotential $W_1$, which in turn determines the SUSY partner potential $V_{1}$ through equation (\ref{susyv1}). We know in advance that in the constant mass case a solution to such equation was $\frac{-\hbar u'}{\sqrt{2m}u}$; hence, in the non constant situation $W_1$ should be somehow related with this expression.

Let us assume that we know a formal eigenfunction $u_1$ of $H_{0}$ with eigenvalue $\epsilon_{1}$,
\be
(H_{0}-\epsilon_{1})u_1=A_1^{\dagger}A_1 u_1=0.
\label{ee45}
\ee
Although at the begining this is a second order differential equation for $u_1$, the operator $A_1$ helps to reduce this order and immediately supplies one solution for $W_1$ since
\be
A_1u_1=\frac{\hbar}{\sqrt{2m(x)}}\frac{d u_1}{dx}+W_1(x) u_1=0.
\ee 
This expression leads to the following $W_1$:
\be
W_1=-\frac{\hbar}{\sqrt{2m(x)}}\frac{u_1'}{u_1}.
\label{sp47}
\ee
In terms of the seed solution $u_1$, the SUSY partner potential $V_1$ can be written as
\be
V_1=V_{0}-\hbar^2 m^{-\frac{1}{2}}\left\{m^{-\frac{1}{2}}\left[\log{(m^{-\frac{1}{4}}u_1)}\right]'\right\}' . \label{v1susy}
\ee
Let us note that we can use any known formal eigenfunction $u_1$ to perform such a transformation. By introducing now equation (\ref{sp47}) into the Riccati equation (\ref{susyv0}), the BDD equation (\ref{ee45}) for $u_1$ is recovered.

Equation (\ref{Inter1}) guarantees that any eigenfunction $\psi_n$ of $H_{0}$ is mapped into an eigenfunction $\phi_{n}$ of $H_{1}$ by acting $A_1$:
\be
\phi_{n} = \tilde{C_n}A_1\psi_n = C_n \, m^{-\frac12} \, u_1^{-1} \, W(u_1,\psi_n), \label{trans-sol}
\ee
where $W(u_1,\psi_n)= u_1\psi_n'-u_1'\psi_n$ is the Wronskian of $u_1$ and $\psi_n$. Let us note that this formula can not supply any eigenfunction of $H_1$ associated to $\epsilon_1$, since $A_1 u_1=0$. However, one can find this so called {\em missing eigenfunction} $\phi_{\epsilon_1}$ of $H_{1}$ associated to $\epsilon_{1}$ by noting that
\be
(H_{1}-\epsilon_{1})\phi_{\epsilon_1} = A_1A_1^{\dagger}\phi_{\epsilon_1} = 0. \label{ext-eq}
\ee
Therefore, one solution to this equation must satisfy:
\be
A_1^{\dagger}\phi_{\epsilon_1} = -\frac{\hbar}{\sqrt{2m(x)}}\frac{d\phi_{\epsilon_1}}{dx}+\left(W_1(x)+\frac{\hbar}{2\sqrt{2}}\frac{m'(x)}{(m(x))^\frac{3}{2}}\right)\phi_{\epsilon_1}=0\Rightarrow
\ee
\be
\frac{\phi_{\epsilon_1}'}{\phi_{\epsilon_1}}=\frac{\sqrt{2m(x)}}{\hbar}\left(W_1(x)+\frac{\hbar}{2\sqrt{2}}\frac{m'(x)}{(m(x))^\frac{3}{2}}\right).
\ee
Introducing now the expression for $W_1$ in terms of the seed solution $u_1$ (see equation (\ref{sp47})) it is obtained
\be
\frac{d\ln{(\phi_{\epsilon_1}  u_1)}}{dx}=\frac{d\ln \sqrt{m}}{dx}\hspace{1cm}\Rightarrow \hspace{1cm}\phi_{\epsilon_1}\propto\frac{\sqrt{m(x)}}{u_1(x)}.
\label{dpg51}
\ee

It is important to note that, if the initial Hamiltonian $H_{0}$ has first-order ladder operators $L^\pm$, then the new Hamiltonian $H_{1}$ will have third-order ladder operators given by 
\be
L_{1}^{\pm}=A_{1}L^{\pm}A_{1}^{+}. \label{3ladder}
\ee

\subsection{Second-order transformations}

In order to implement another first-order transformation we need to choose a formal eigenfunction $v_2$ of $H_1$ associated to the factorization energy $\epsilon_2$; this seed solution is annihilated by the first-order differential operator
\be
A_2=\frac{\hbar}{\sqrt{2m(x)}}\frac{d}{dx}+W_2(x) = \frac{\hbar}{\sqrt{2m(x)}} \left(\frac{d}{dx} - \frac{v_2'}{v_2} \right),
\ee
which, together with $A_2^\dagger$, fulfills the following operator relations:
\bea
\hskip-1cm H_{1}-\epsilon_{2}=A_2^{\dagger}A_2, \hspace{.3cm}H_{2}-\epsilon_{2}=A_2A_2^{\dagger}\hspace{.1cm}\Rightarrow\hspace{.1cm}H_{1}A_2^{\dagger}=A_2^{\dagger}H_{2},\hspace{.3cm} H_{2}A_2=A_2H_{1}.
\label{Inter2}
\eea
Thus, the eigenfunctions $\chi_n(x)$ of $H_2$ can be obtained by acting $A_2$ onto the corresponding ones $\phi_n(x)$ of $H_1$:
\be
\chi_{n} = \tilde{F_n}A_2\phi_n = F_n (m^{-1/2} v_2^{-1}) W(v_2,\phi_n). \label{trans2-sol}
\ee

The expression for the new potential $V_2$ is determined by the form taken by $v_2$, which in turn depends on the value taken by the factorization energy $\epsilon_2$. We can distinguish two different cases.

\subsubsection{Non-confluent case with $\epsilon_2\neq \epsilon_1$.}
Let us suppose first that $\epsilon_2\neq \epsilon_1$, so that the seed solution $v_2$ employed in the second first-order transformation is obtained by acting $A_1$ on the corresponding formal eigenfunction $u_2$ of $H_0$ associated to $\epsilon_2$ as follows (see also equation (\ref{trans-sol})):
\be
v_2 \propto A_1 u_2 \propto m^{-\frac12} u_1^{-1} W(u_1,u_2).
\ee
Thus, the new potential becomes (see equation (\ref{v1susy})):
\bea
V_2 & = & V_{1}-\hbar^2 m^{-\frac{1}{2}}\left\{m^{-\frac{1}{2}}\left[\log{(m^{-\frac{1}{4}}v_2)}\right]'\right\}' \nonumber \\
& = & V_0 - \hbar^2 m^{-\frac{1}{2}}\left\{m^{-\frac{1}{2}}\left[\log{(m^{-\frac12} u_1 v_2)}\right]'\right\}' \nonumber \\
& = & V_0 - \hbar^2 m^{-\frac{1}{2}}\left\{m^{-\frac{1}{2}}\left[\log{(m^{-1} W(u_1,u_2))}\right]'\right\}'.
\eea
The eigenfunctions of $H_2$ are obtained, in general, from equation (\ref{trans2-sol}).
However, once again there is a formal eigenfunction $\chi_{\epsilon_2}$ of $H_2$ associated to $\epsilon_2$, annihilated by $A_{2}^{\dagger}$, which is given by:
\be
\chi_{\epsilon_2}\propto \frac{\sqrt{m(x)}}{v_2(x)}\propto \frac{m(x)u_1(x)}{W(u_1,u_2)}.
\ee

Let us note that there exist now a pair of second-order differential operators $A_2 A_1, \ A_1^\dagger A_2^\dagger$, intertwining $H_0$ and $H_2$ as follows:
\be
H_2 \, (A_2A_1) = (A_2 A_1) \, H_0, \qquad H_0 \, (A_1^\dagger A_2^\dagger) = (A_1^\dagger A_2^\dagger) \, H_2. \label{2inter}
\ee
Moreover, if the initial Hamiltonian $H_{0}$ has first-order ladder operators $L^\pm$, then the new Hamiltonian $H_{2}$ will have fifth-order ladder operators given by 
\be
L_{2}^{\pm}=(A_{2}A_{1})\,L^{\pm}\,(A_{1}^{+}A_{2}^{+}). \label{5ladder}
\ee

\subsubsection{Confluent case with $\epsilon_2 = \epsilon_1$.} Let us suppose now that the seed solution $v_2$ used to implement the second transformation has a factorization energy equal to the one employed in the first transformations, i.e., $\epsilon_2 = \epsilon_1$. Moreover, we are going to take as $v_2$ the general solution to the stationary Schr\"odinger equation for $H_1$ associated to $\epsilon_1$. Since we already know one solution to this equation (see Eqs.~(\ref{ext-eq},\ref{dpg51})), we can use Abel's identity \cite{Arfken} to find such a $v_2$:
\bea
&& v_2(x) = \frac{\sqrt{m(x)}}{u_1(x)}w(x),
\eea
where
\bea
&& w(x) = \left((1-d)+d\int u_1^{2}(x)dx\right),\hspace{1cm}d\in[0,1].
\eea
Thus, the new potential in this case becomes:
\bea
V_2 & = & V_{1}-\hbar^2 m^{-\frac{1}{2}}\left\{m^{-\frac{1}{2}}\left[\log{(m^{-\frac{1}{4}}v_2)}\right]'\right\}' \nonumber \\
& = & V_0 - \hbar^2 m^{-\frac{1}{2}}\left\{m^{-\frac{1}{2}}\left[\log{(m^{-\frac12} u_1 v_2)}\right]'\right\}' \nonumber \\ 
& = & V_0 - \hbar^2 m^{-\frac{1}{2}}\left\{m^{-\frac{1}{2}}\left[\log(w)\right]'\right\}'. \label{conflmx}
\eea 
The formal eigenfunction $\chi_{\epsilon_2}$ of $H_2$ associated to $\epsilon_2=\epsilon_1$ is now given by: 
\be
\chi_{\epsilon_2}\propto \frac{\sqrt{m(x)}}{v_{2}(x)}\propto \frac{u_1(x)}{w(x)} .
\ee

Let us stress that equation (\ref{conflmx}) generalizes the corresponding confluent formula for the constant mass situation \cite{Fer2003}. In addition, once again there are two second-order intertwining operators satisfying equation (\ref{2inter}) and two fifth-order ladder operators given by equation (\ref{5ladder}), provided that $H_0$ has first-order ladder operators $L^\pm$.

%%%%%%%%%%%%%%%%%%%%%%%%%%%%%%%%%%%%%%%%%%%%%
\section{SUSY transformations for the cosine mass profile}

\subsection{First order SUSY transformation}
Let us take the first excited state $\psi_{1}(x)$ of example \ref{cosproex} as the seed solution $u_1(x)$ to perform a first-order transformation, which induces a singularity at $x=0$ so that the modified domain can be though of as being $(0,\infty)$, as it was done in \cite{Fer2014}.

\begin{figure}[htp]
\centering
\includegraphics[height=7cm, width=12cm]{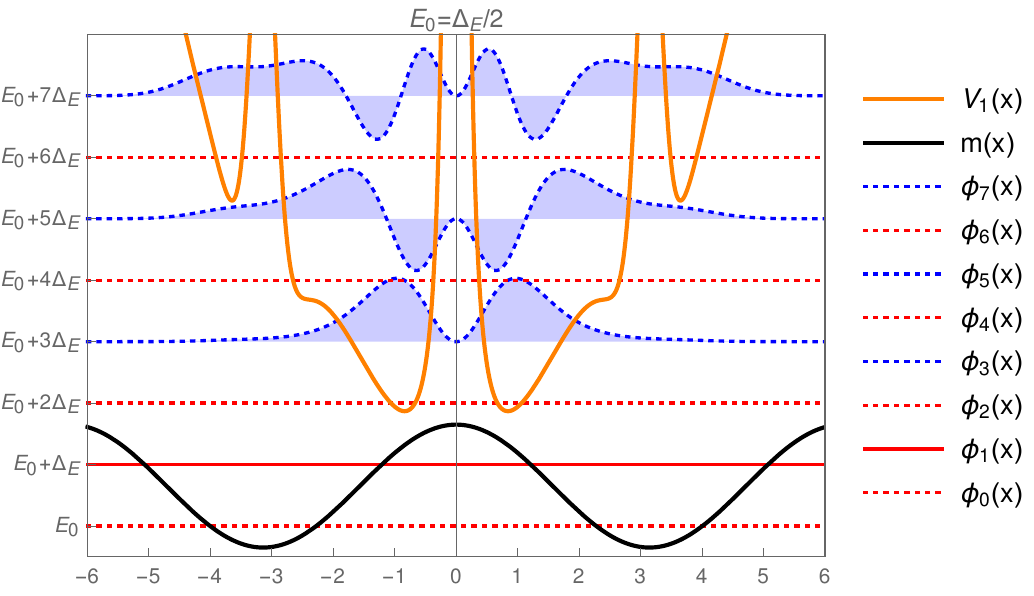}
\caption{Singular first-order SUSY transformation employing the first excited state of $H_{0}$. The domain of the SUSY partner potential $V_{1}$ becomes $(0,\infty)$; the wavefunctions in red are singular at $x=0$, with the exception of $\phi_{1}$ which is the null vector.}
\label{fig5}
\end{figure}

The superpotential is thus given by
\bea
W_{1}\!&=&\!\Delta_E \sqrt{2}\sqrt{m_0+1} E\left[\frac{x}{2},\frac{2}{m_0+1}\right]-\frac{1}{2 \sqrt{2} \sqrt{m_0+1} E\left[\frac{x}{2},\frac{2}{m_0+1}\right]}\nonumber\\
&&+\frac{\sin (x)}{4 \sqrt{2} \left(m_0+\cos (x)\right){}^{3/2}}.
\label{W1}
\eea
The SUSY partner potential of $V_0$ is easily calculated, by inserting equations (\ref{W1}) and (\ref{Vcos}) into equation (\ref{susyv1}) (the shape of the potential and its first eigenfunctions are illustrated in Figure \ref{fig5}), leading to:
\bea
V_{1}&=&\Delta_E+ 2 \Delta_E^2 (m_0\!+\!1) E\left[\frac{x}{2},\frac{2}{m_0+1}\right]^2\!+\!\frac{3 \cos ^2(x)\!-\!7\!-\!4 m_0 \cos (x)}{32 (m_0+\cos (x))^3} \nonumber
\\ &&  +\frac{1}{4 (m_0+1) E\left[\frac{x}{2},\frac{2}{m_0+1}\right]^2}. 
\eea

\subsection{Non-confluent second-order transformation}

\begin{figure}[htp]
\centering
\includegraphics[height=7cm, width=12cm]{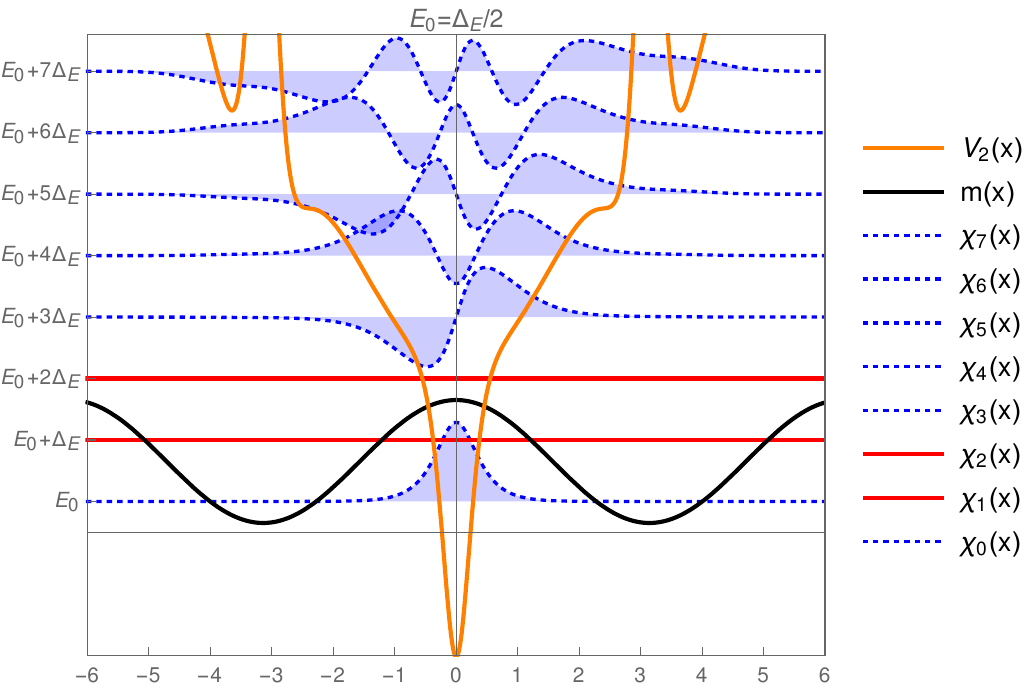}
\caption{Plot of the second-order SUSY partner potential $V_{2}$ and the corresponding eigenfunctions of $H_{2}$. The eigenfunctions $\chi_{1}$ and $\chi_{2}$ become identically zero.}
\label{fig6}
\end{figure}

The formal eigenfunction $\phi_{2}$ has an eigenvalue ($5\Delta_E/2$) which does not belong to the spectrum of $H_{1}$. Let us perform now the second first-order transformation employing $v_2=\phi_{2}$. 
The new superpotential is thus given by  
\bea
\hspace{-2.5cm}W_{2}&&=\frac{\sin (x)}{4 \sqrt{2} \left(m_0+\cos (x)\right)^{3/2}}+\frac{8 \sqrt{2} \Delta _{E}^2 \left(\left(m_0+1\right) \left(m_0+\cos (x)\right)\right)^{3/2} E\left[\frac{x}{2},\frac{2}{m_0+1}\right]^3}{\left(m_0+\cos (x)\right)^{3/2} \left(1+8\Delta _{E} \left(m_0+1\right)E\left[\frac{x}{2},\frac{2}{m_0+1}\right]^2\right)}
\nonumber
\\
\hspace{-2.5cm}&&+\frac{1-4\Delta _{E} \left(m_0+1\right) E\left[\frac{x}{2},\frac{2}{m_0+1}\right]^2}{2 \sqrt{2} \sqrt{m_0+1} E\left[\frac{x}{2},\frac{2}{m_0+1}\right] \left(1+8\Delta _{E} \left(m_0+1\right) E\left[\frac{x}{2},\frac{2}{m_0+1}\right]^2\right)}.
\label{sup2}
\eea
It is interesting to see that, after this second step, the singularity that was created at $x=0$ for $V_{1}$ is now removed from $V_2$ (see Figure \ref{fig6}). The new SUSY partner potential $V_{2}(x)$ becomes
\bea
\hspace{-2.5cm}V_{2}=&2 \Delta _{E}^2\left(m_0+1\right) E\left[\frac{x}{2},\frac{2}{m_0+1}\right]{}^2+\frac{-8 m_0 \cos (x)+3 \cos (2 x)-11}{64 \left(m_0+\cos (x)\right){}^3}\nonumber\\
\hspace{-2.5cm}&+\frac{-2 \Delta _{E}+64 \Delta _{E}^2\left(m_0+1\right)  E\left[\frac{x}{2},\frac{2}{m_0+1}\right]{}^2 \left(1+2 \Delta _{E}\left(m_0+1\right)  E\left[\frac{x}{2},\frac{2}{m_0+1}\right]{}^2\right)}{\left(1+8\Delta _{E} \left(m_0+1\right) E\left[\frac{x}{2},\frac{2}{m_0+1}\right]{}^2\right){}^2}.
\eea
Notice that now it appears a gap in the spectrum of $H_2$, namely, Sp$H_2 =\{\Delta_E/2,7\Delta_E/2,9\Delta_E/2,\dots \}$.

\subsection{Confluent second-order SUSY transformation}

\begin{figure}[htp]
\centering
\includegraphics[height=7cm, width=12cm]{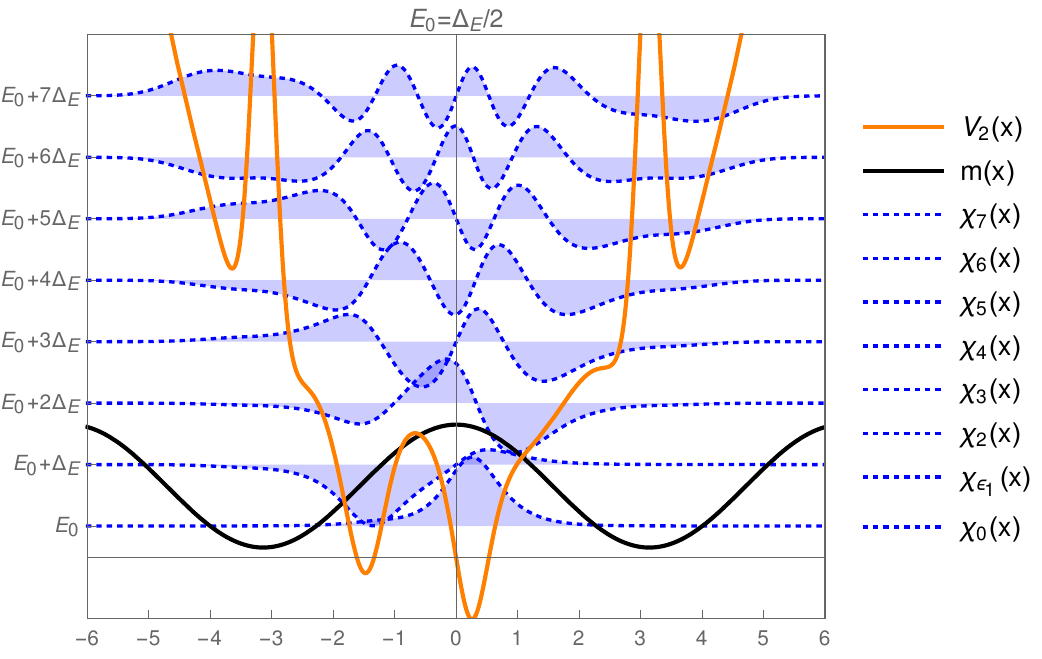}
\caption{Confluent second-order SUSY partner potential (Eq.~(\ref{conflmx})) for the cosine mass profile when the seed solution is the first excited state of $H_0$. This transformation is nonsingular for $d<0.360691$.}
\label{fig7}
\end{figure}

Instead of taking $\phi_{2}$ as the seed solution to implement the second transformation, let us employ now the formal eigenfunction (\ref{dpg51}) of $H_{1}$ associated to $\epsilon_{1}$ to address the confluent algorithm \cite{Fer2003}. Note that, in general, the specific conditions to produce a nonsingular transformation in this case are still missing, although for constant mass problems it is well known that the use of an eigenfunction of the initial Hamiltonian can produce non-singular confluent transformations. An example of this type of potentials, generated through the confluent algorithm, is shown in Figure \ref{fig7}.

\newpage

\section{Concluding remarks}

In this paper we have started the analysis of a general problem, the determination of position dependent mass systems ruled by polynomial Heisenberg algebras and the study of their SUSY partners. In order to implement such program, it is natural to address first the case with the lowest degree ($m=0$) and to proceed later by increasing gradually such a $m$. 

We have presented here precisely the way to determine the general position dependent mass systems which have first order ladder operators and thus an infinite equidistant spectrum, by demanding that the zero degree polynomial Heisenberg algebra rules the corresponding Ben-Daniel-Duke Hamiltonian, without attempting to establish a connection with a constant mass problem. Once these systems have been determined, we have applied to them the quantum mechanical SUSY treatment to generate the corresponding SUSY partners of first and second order (with the same mass profile), and we have identified their natural ladder operators, which turn out to be of order greater than one. We have found also compact formulas to calculate the SUSY partner potentials of first and second-order. As far as we know, for systems ruled by BDD Hamiltonians the second order expressions are new. We have shown also that it is possible that the associated SUSY partners can have spectral gaps, when compared with the equally spaced initial spectrum. This suggests that there are wide possibilities to implement the spectral design through SUSY QM, when applied to position dependent mass systems. We believe that this subject is worth addressing more deeply in the near future.

\section{Acknowledgments}

MIED Acknowledges the support of Conacyt, grant 489860.

%%%%%%%%%%%%%%%%%%%%%%%%%%%%%%%%%%%%%%%%%%%%
\section*{References}
%\bibliography{Bibliografia.bib}{}
%\bibliographystyle{unsrt}

\end{document}